\documentclass[pre,aps,showpacs,preprint,amsmath,letterpaper,endfloats]{revtex4}
\usepackage{graphicx}
\usepackage{mathrsfs}
\usepackage{wasysym}
\usepackage{textcomp}
\usepackage{pifont}

\begin{document}

\title{Evanescent wave interference and the total transparency of a warm
high-density plasma slab}
\author{E. Fourkal$^{1}$}
\author{I. Velchev$^{1}$}
\author{C-M. Ma$^{1}$}
\author{A. Smolyakov$^{2}$}

\affiliation{$^{1}$Department of Radiation Physics, Fox Chase Cancer Center,
Philadelphia, PA 19111, U.S.A.} 
\affiliation{$^{2}$Department of Physics and
Engineering Physics, University of Saskatchewan, Saskatoon, Canada}

\begin {abstract}
It is shown that an overcritical density plasma slab can be totally
transparent to a \textit{p}-polarized obliquely incident electromagnetic
wave. High transparency is achieved due to the interference of the
evanescent waves in the subcritical region. The transmission coefficient has
the resonant character due to the excitation of a plasma surface mode
(plasmon-polariton). It is crucial that two evanescent waves have certain
phase shift to provide non-zero energy flux through non-transparent region.
The required phase shift is obtained at the exact resonance and corresponds
to the absolute transparency. In a warm plasma case, the excitation of the
propagating longitudinal (electrostatic) modes becomes possible. We
demonstrate that these longitudinal excitations facilitate the total
transparency of an opaque plasma slab creating additional resonances in the transmission property of the system.
\end {abstract}

\date{\today}

\maketitle

\section{Introduction}

There exists a large number of theoretical and experimental studies
dedicated to the problem of light interaction with dense media having
negative dielectric permittivity~\cite
{kindel75,aliev77,dragila85,bliokh05,fang05}. The increasing interest in the
properties of such media has been driven by their potential applications in
different branches of science and technology. Understanding of the
fundamental processes that take place during the interaction of
electromagnetic radiation with overcritical density plasmas ($\epsilon
=1-\omega _{p}^{2}/\omega ^{2}<0$ for $\omega <\omega _{p}$) is of great
importance to a number of areas. In particular, one of the challenges of the
inertial confinement fusion (ICF) experiments is the effective heating of
the fuel pellet by the laser light (in the direct drive) or X-rays (in the
indirect drive ignition experiments). The two well known absorption
mechanisms in plasma are of collisional~\cite{dawson62,langdon80} and
resonant~\cite{denisov57,freidberg72,forslund75} nature. In the collisional
absorption process, the energy of light is converted into thermal electron
motion through electron-ion collisions. In the resonant absorption
mechanism, the linear conversion in nonuniform plasma leads to a resonant
excitation of electrostatic plasma wave, which subsequently decays giving
rise to energetic electrons. 

In this work we investigate another resonant mechanism that dramatically
enhances the penetration of the electromagnetic field into overdense plasma.
This mechanism involves surface plasma wave or plasmon-polariton~\cite
{trivelpiece59,kaw70} which may be excited at the plasma-plasma interface.
Unlike bulk plasma waves, which are electrostatic in nature, the surface
waves are of mixed longitudinal-transverse type and can couple to radiating
electromagnetic fields. As was shown in Ref.~\cite{aliev77}, at certain
conditions the energy of the incident electromagnetic radiation can be
completely absorbed by the plasma through the resonant excitation of surface
plasma waves which subsequently dissipate through standard irreversible
processes like linear wave transformation or collisions.

In this paper we show that the resonant excitation of surface modes may
create the condition for absolute transparency of an overdense plasma slab.
The transparency mechanism studied in our work is directly related to the
problem of the anomalous transparency of light in materials with negative
permittivities. Media with negative electric permittivity and magnetic
permeability~\cite{veselago68} (the so-called left-handed materials or LHM)
possess some unusual optical properties, in particular, the possibility of
creating an ''ideal lens'' offering sub-wavelength resolution~\cite
{pendry00,fang05,podolskiy05}. As was pointed out by Pendry~\cite{pendry00},
the superlens is essentially based on the possibility of evanescent waves
amplification, facilitated by the excitation of surface plasmons. The
amplification of the evanescent waves and their interference are the basic
mechanisms behind the anomalous transparency demonstrated in our work. This
indicates the feasibility of superlensing in materials without any
restrictions on the magnetic permeability $\mu $, which has also been noted
by Pendry.

In Ref.~\cite{dragila86}, the authors suggested that the incident
electromagnetic wave may excite a pair of coupled surface plasma waves on
both sides of the slab. The energy of the surface wave running on the
opposite side of the slab can be dissipated by reemission of the
electromagnetic waves, which causes the layer to appear totally transparent.
However a special symmetric structure consisting of media with different
dielectric permittivities has to be formed in order for a pair of coupled
surface waves to exist.

In this paper we reexamine the conditions for total transparency of a layer
of overcritical density plasma. Anomalously high transmission of light is
explained through the  interference of the evanescent fields formed by the
incident/emitted electromagnetic radiation and the  field of the surface
plasma wave. We show that the total transparency of light can be achieved
even when a \emph{single} surface plasma mode is excited. Therefore, there
is no need to form a symmetric structure in which an overcritical density
plasma slab is sandwiched between two additional layers with specially
chosen dielectric permittivities. In other words anomalous light
transmission can be achieved by passing the light through a single
plasma-plasma interface. In most laboratory experiments when high-power
laser light interacts with a solid density target, a prepulse precedes the
main laser pulse, creating the plasma density profiles discussed in this
paper.

\section{Surface wave assisted total light transmission in cold plasmas}

As it was noted above, high transparency of dense plasma occurs as a result
of the resonant excitation of a surface mode. Since the surface of the
medium with $\epsilon<0$ behaves like surface wave resonator \cite{bliokh05}%
, the problem of surface wave excitation is equivalent to that of resonator
excitation. Under resonant conditions (external field frequency and its wave
vector's tangential component have to match those determined from the
dispersion relation for the plasma surface wave) the coupling between the
surface eigenmode and the external field becomes extremely efficient. The
minimum in the reflection coefficient (the maximum in the transmission
coefficient in the absence of dissipation) occurs for these resonant values
of the external wave frequency and wave number.

Here we review a dispersion relation for surface waves that exist on a cold
plasma-plasma interface for arbitrary values of dielectric permittivities of
the adjacent regions. It is found by solving the linearized set of cold
fluid-Maxwell equations, 
\begin{subequations}
\label{eqs1}
\begin{eqnarray}
\mathbf{\nabla \times E}=-\frac{1}{c}\frac{\partial \mathbf{B}}{\partial t} \label{eq1} \\
\mathbf{\nabla \times B}=-\frac{4\pi n_0
e}{c}\mathbf{\tilde{v}}+\frac{1}{c}\frac{\partial
\mathbf{E}}{\partial t} \label{eq2}\\
\frac{\partial \mathbf{\tilde{v}}}{\partial
t}=-\frac{e}{m}\mathbf{E},
\label{eq3}
\end{eqnarray}
\end{subequations}
where ions are assumed to be motionless. Letting all physical quantities
vary as $\mathbf{\Psi}(x)\exp\left({ik_yy-i\omega t}\right)$, where $x$ is
the coordinate into the plasma and $y$ runs along the interface with the
field components of the form $\mathbf{E}$=$(E_x,E_y,0)$, $\mathbf{B}$=$%
(0,0,B_z)$ one obtains, 
\begin{equation}  \label{eqs2}
\epsilon \frac{d}{dx}\left(\frac{1}{\epsilon}\frac{dB_z}{dx}%
\right)-\alpha^2B_z=0,
\end{equation}
where $\alpha^2=\left(k^2_y-\frac{\omega^2}{c^2}\epsilon\right)$ and $%
\epsilon=1-\omega^2_p/\omega^2$. Assuming that the interface is located at
the plain $x=0$, and for $x<0$ plasma has a constant density such that $%
\omega_p=\omega_{p1}$, and for $x>0$ a different constant density $%
\omega_p=\omega_{p2}$, the solution to Eq.~(\ref{eqs2}) in both regions is, 
\begin{eqnarray}
B_z=A_1\exp{(\alpha_1x)}, & x<0  \nonumber \\
B_z=A_2\exp{(-\alpha_2x)}, & x>0  \nonumber
\end{eqnarray}
It should be noted that here we only keep those solutions that decay away
from the interface (a condition for existence of a well-defined surface
mode). Using the boundary conditions at $x=0$ requiring the continuity of $%
B_z$ and $dB_z/dx/\epsilon$, one readily obtains the dispersion relation for
the surface waves on the cold plasma-plasma interface, 
\begin{equation}  \label{eqs3}
\alpha_1/\epsilon_1+\alpha_2/\epsilon_2=0\rightarrow k_y=\frac{\sqrt{%
\left(\Delta-\omega^2\right)\left(\omega^2-1\right)}} {\sqrt{%
1+\Delta-2\omega^2}}
\end{equation}
where $\Delta=\omega^2_{p1}/\omega^2_{p2}$, $\omega$ is normalized to the
plasma frequency in the region 2 where $\epsilon<0$ and $k_y$ is normalized
to the classical skin depth $\delta=c/\omega_{p2}$. As can be seen from Eq.~(%
\ref{eqs3}), $\epsilon_1$ and $\epsilon_2$ must have opposite signs, so that
one plasma region must have undercritical electron density while the
other-overcritical.

One important difference between the surface plasmons running on a
plasma-vacuum interface  (obtained from (\ref{eqs3}) by setting $\Delta=0$))
and those on a plasma-plasma  interface lies in a fact that the phase
velocity in the latter case can exceed that of  light, whereas the phase
velocity of the former is always subluminal.  Because of this fact, the
surface plasmons on a plasma-plasma interface can couple  to radiating
electromagnetic fields, which are evanescent in plasma but may become 
propagating if a vacuum region is present beyond plasma layers. The opposite
is also true: if an external $p$-polarized electromagnetic pulse  is
obliquely incident on the structure comprising two plasma layers, it may
under a resonant condition excite a surface plasma mode. On the contrary, a $%
p$-polarized electromagnetic pulse obliquely incident on a vacuum-plasma
interface  cannot excite a plasma surface wave since the phase velocity of
this collective mode  can never exceed the speed of light (the perturbation
of the plasma surface by the  external field propagates in this case with
phase velocity greater than that of light)  and the field of the external
radiation is not evanescent (as we mentioned earlier,  the fields of the
surface wave are decaying away from the interface, or only  evanescent
external fields can excite the plasma surface mode). This means that  if a
layer of dielectric is added, so that a system comprising dielectric medium,
vacuum and overcritical density plasma is formed in such a way that the
external  field in vacuum is evanescent (through total internal reflection
of light in  dielectric material) and the perturbation of the plasma surface
by the evanescent  field propagates with the phase velocity that is lower
than the speed of light,  then the external electromagnetic field may excite
the surface plasma mode on a  vacuum-plasma interface~\cite
{dragila85,bliokh05}.

\begin{figure}[t]
\centering
\centerline{\includegraphics[width=0.9\columnwidth]{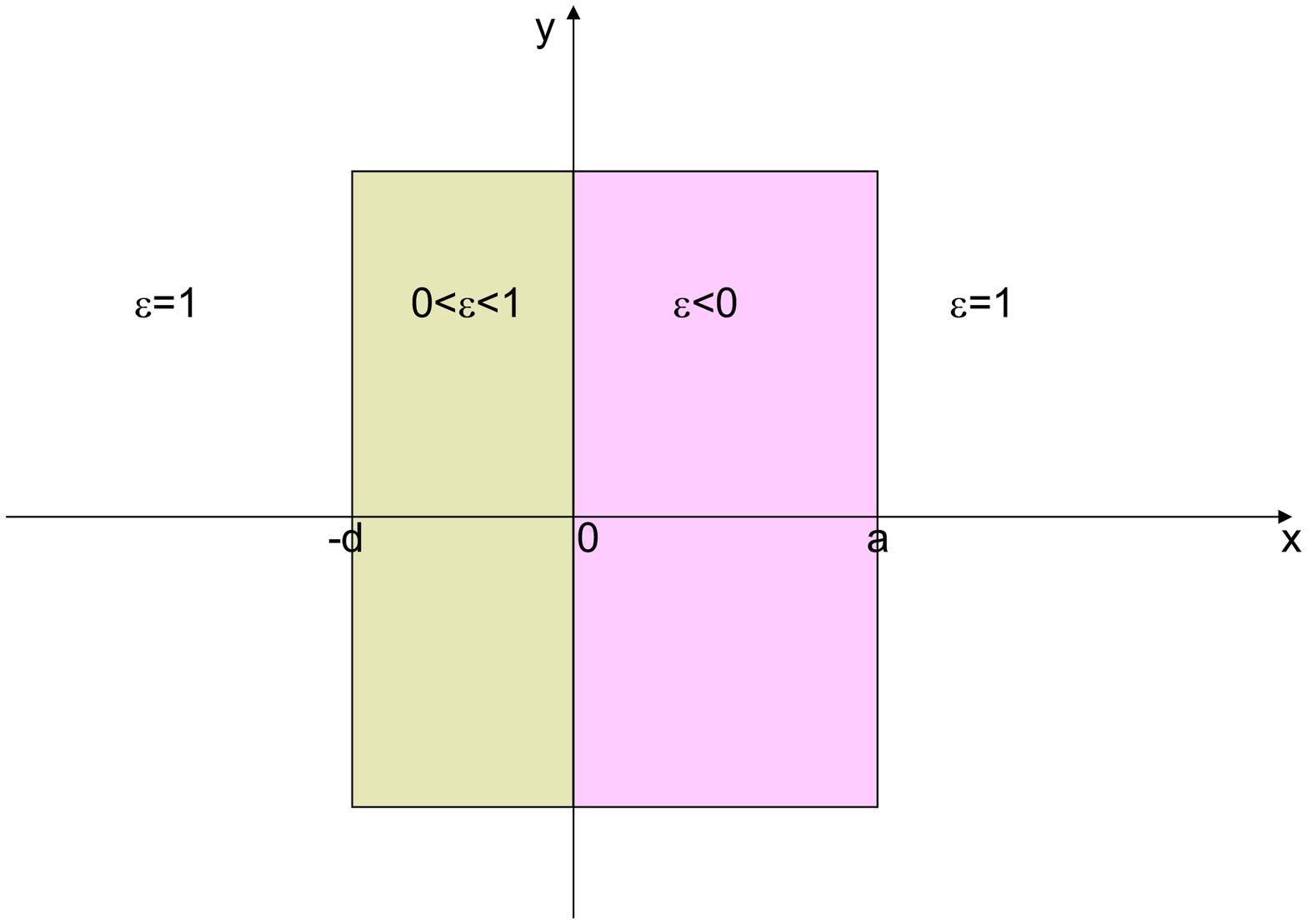}}
\caption{Schematic diagram of the spatial density (dielectric constant)
distribution.}
\label{fig1}
\end{figure}

Optical properties of a double-plasma system are determined by matching the
fields of the incident/reflected electromagnetic waves with those of the
surface wave. Let us consider a system which is characterized by spatial
density  (dielectric constant) distribution shown in Fig.~(\ref{fig1}). A $p$
-polarized plane  electromagnetic wave is obliquely incident on this
structure. In both plasma regions,  the electromagnetic field $\mathbf{B}%
=(0,0,B_z)$ is the solution to the following  equation: 
\begin{equation}
\epsilon \frac{d}{dx}\left(\frac{1}{\epsilon}\frac{dB_z}{dx}\right)+\frac{%
\omega^2}{c^2}\left(\epsilon - \sin^2\theta\right)B_z=0,  \label{eqs4}
\end{equation}
with a general solution having the following from, 
\begin{equation}
B_z=A_1e^{ikx}+A_2e^{-ikx}  \label{eqs5}
\end{equation}
where $k=k_0\sqrt{(\epsilon-\sin^2~\theta)}$, $k_0=\omega/c$. The
electromagnetic fields in the vacuum regions ($x<-d$ and $x>a$) have the
following form, 
\begin{eqnarray}
B_z=E_0e^{ik_0\cos \theta~x}+Re^{-ik_0\cos \theta~x},& x<-d  \nonumber \\
B_z=Te^{ik_0\cos \theta~x},& x>a  \nonumber \\
\nonumber
\end{eqnarray}
where $T$ and $R$ designate the field components for the transmitted and
reflected waves. Matching solutions at different boundaries by requiring
continuity of $B_z$ and $dB_z/dx/\epsilon$ across interfaces, one obtains a
system of six equations (not shown here) for the unknown coefficients. The
expression for the transmission coefficient has the following form, 
\begin{eqnarray}  \label{eqs6}
&T=\left|\frac{4e^{i(k_1d+k_2a-(a+d)k_0\cos\theta)}k_0k_1k_2\epsilon_1%
\epsilon_2\cos\theta}{\left(\left(1+e^{2ik_1d}\right) k_0k_1k_2\epsilon_1
\epsilon_2\cos\theta+e^{2ik_2a}\left(1+e^{2ik_1d}\right)k_0k_1k_2\epsilon_1
\epsilon_2\cos\theta-2ie^{i(k_1d+k_2a)}\Upsilon \right)}\right|^2 \\
&\Upsilon=k_2\epsilon_2\cos[k_2a]\left(k^2_1+k^2_0\epsilon^2_1\cos^2\theta
\right)\sin[k_1d]+  \nonumber \\
&\left(k_1\epsilon_1\cos[k_1d]\left(k^2_2+k^2_0\epsilon^2_2\cos^2\theta
\right)- ik_0\left(k^2_2\epsilon^2_1+k^2_1\epsilon^2_2\right)\cos\theta \sin[%
k_1d]\right)\sin[k_2a],  \nonumber
\end{eqnarray}
where $k_{1,2}=k_0\sqrt{(\epsilon_{1,2}-\sin^2\theta)}$. For resonant values
of the external electromagnetic field frequency and incidence angle, one
should expect anomalously high transmission of light.

\begin{figure}[t]
\centering
\centerline{\includegraphics[width=0.9\columnwidth]{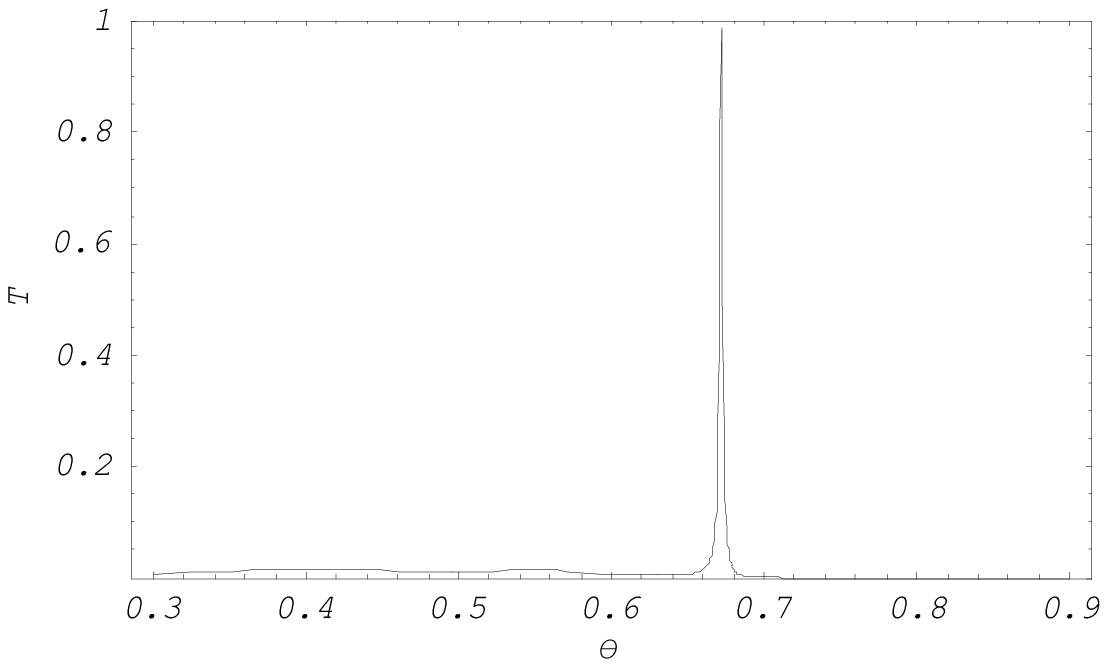}}
\caption{Transmission coefficient as a function of incidence angle. A
double-layer plasma system becomes completely transparent at the incidence
angle $\theta$=0.671955. Undercritical and overcritical plasma slabs
dielectric constants and thicknesses are $\epsilon_1=0.3428$, $%
d=27*c/\omega_{p2}$ and $\epsilon_2$=-2.97, $a=3.12*c/\omega_{p2}$
correspondingly. External radiation frequency $\omega/\omega_{p2}=0.5019$. }
\label{fig2}
\end{figure}

Figure (\ref{fig2}) shows the transmission coefficient as a function of the
incidence angle. As one can see, there is a sharp increase in the
transmission properties of the system when the angle of incidence matches
one of many resonant values of $\theta=0.671955$ radians at which point the
transmission coefficient reaches the value $T=0.99999$. The resonant values
exactly correspond to those given by an expression (\ref{eqs3}) for the
dispersion relation of the surface plasma wave. Thus, the anomalous
transmission occurs even for the system consisting of an undercritical
density plasma  layer adjacent to an overcritical density plasma slab, so
that there is no need to  form a sandwich-like structure as argued in Refs. (%
\cite{dragila85,dragila87,bliokh05})  in which a second surface plasma wave
has to be excited on the opposite side of the  overcritical density layer to
achieve the same effect. In other words, the anomalous  light transmission
can be achieved through the excitation of a single surface plasma wave.  It
is interesting to note that for a sandwich-like system, there are generally
two  peaks in the transmission coefficient at two different but close angles
of  incidence\cite{dragila86}. When the second layer of undercritical plasma
is  removed, or the thickness of the overcritical density plasma slab
becomes too  large, the two solutions degenerate into one, which corresponds
to the surface  plasmon running on a single plasma-plasma interface.
Analogous effect is seen in  quantum mechanics where the energy level of a
particle in two-humped potential is  split into two energy states that are
close to each other\cite{landau_q}. As the  potential is transformed into a
single well, the two energy levels degenerate  into a single energy state.

\begin{figure}[t]
\centering
\centerline{\includegraphics[width=0.9\columnwidth]{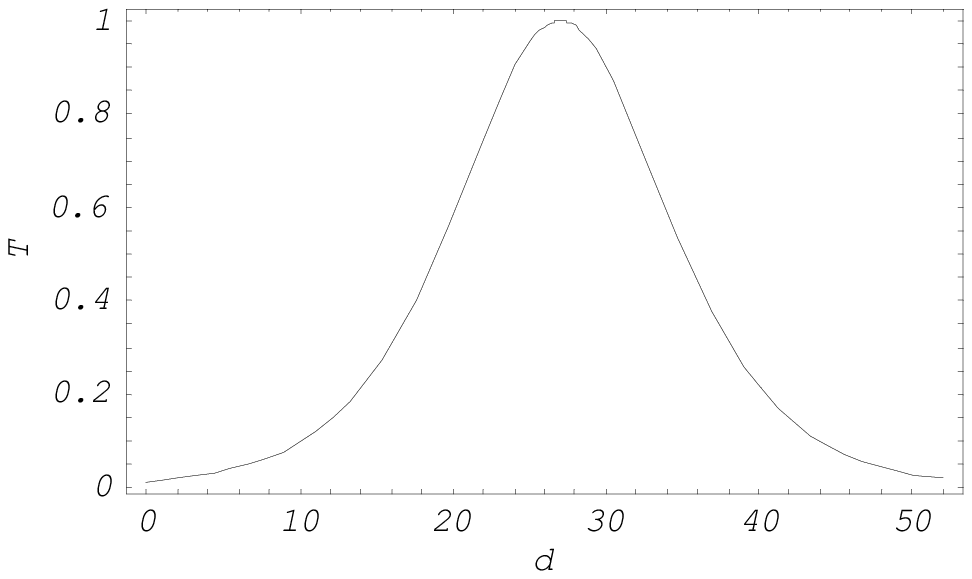}}
\caption{Transmission coefficient as a function of a underdense plasma layer
thickness. Overcritical plasma slab dielectric constant and thickness are $%
\epsilon_2$=-2.97, $a=3.12*c/\omega_{p2}$ correspondingly. Underdense plasma
layer dielectric constant is $\epsilon_1=0.3428$. External radiation
frequency $\omega/\omega_{p2}=0.5019$. Incidence angle $\theta=0.671955$.}
\label{fig3}
\end{figure}

\begin{figure}[t]
\centering
\centerline{\includegraphics[width=0.9\columnwidth]{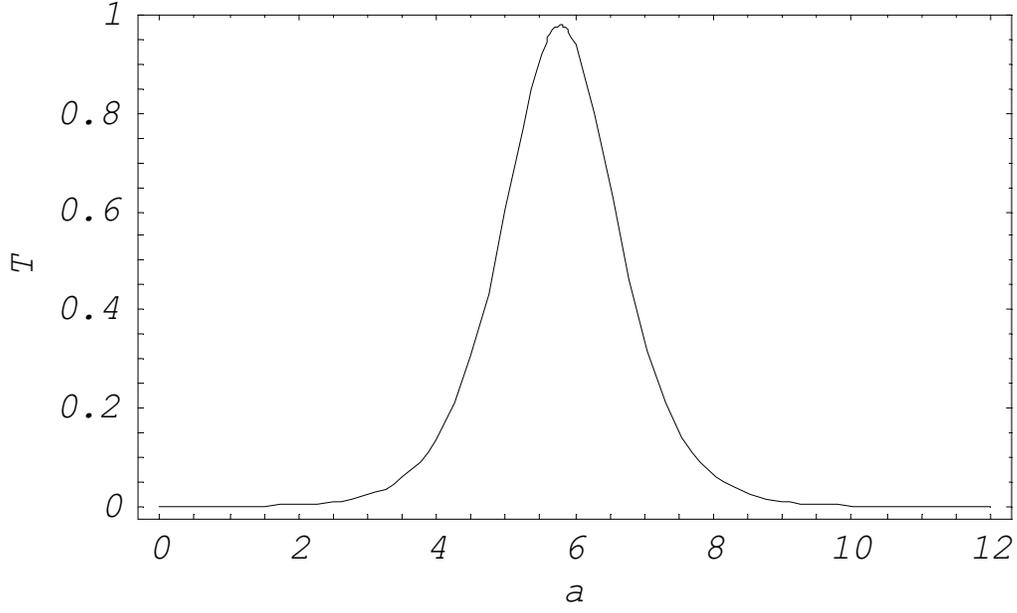}}
\caption{Transmission coefficient as a function of an overdense plasma layer
thickness. Overcritical plasma slab dielectric constant is $\epsilon_2$%
=-2.97. Underdense plasma layer dielectric constant and thickness are $%
\epsilon_1=0.3428$ and $d=50*c/\omega_{p2}$ correspondingly. External
radiation frequency $\omega/\omega_{p2}=0.5019$. Incidence angle $%
\theta=0.671955$.}
\label{fig4}
\end{figure}

\begin{figure}[t]
\centering
\centerline{\includegraphics[width=0.9\columnwidth]{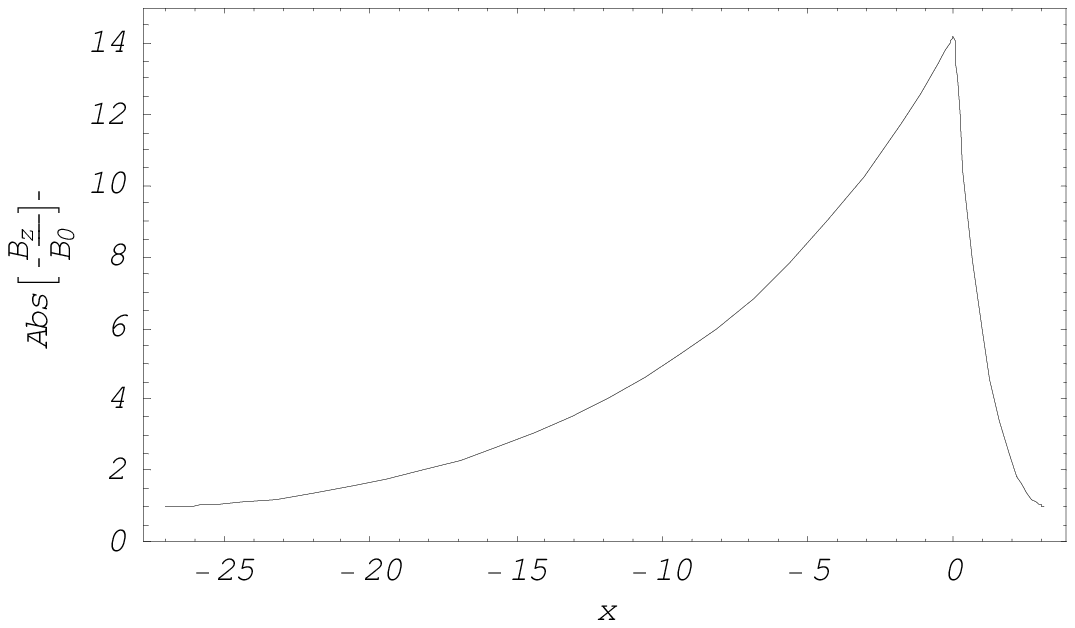}}
\caption{Spatial distribution of an absolute value of the magnetic field
associated with a single surface wave present on a plasma-plasma boundary.
Undercritical and overcritical plasma slabs dielectric constants and
thicknesses are $\epsilon_1=0.3428$, $d=27*c/\omega_{p2}$ and $\epsilon_2$%
=-2.97, $a=3.12*c/\omega_{p2}$ correspondingly. External radiation frequency
and angle of incidence are $\omega/\omega_{p2}=0.5019$, $\theta=0.671955$
correspondingly. $B_0$ is the magnitude of the magnetic field in the
incident wave.}
\label{fig5}
\end{figure}

\begin{figure}[t]
\centering
\centerline{\includegraphics[width=0.9\columnwidth]{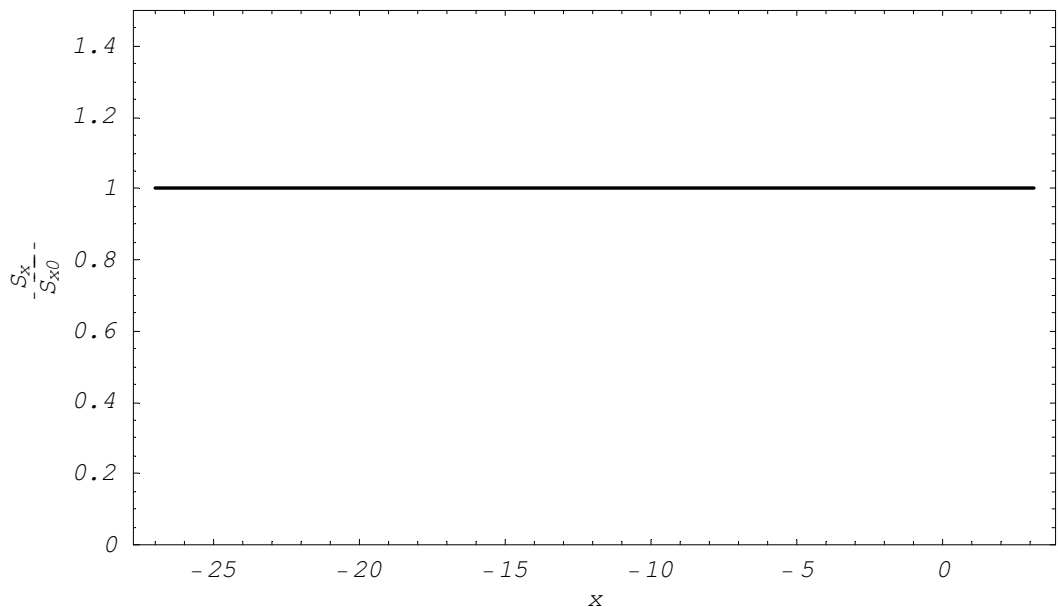}}
\caption{Spatial distribution of the $x$-component of the Poynting vector $%
S_x$, normalized to the $x$-component of the Poynting vector of the incident
wave $S_{x0}=1/2E^2_0\cos\theta$.}
\label{fig6}
\end{figure}

\begin{figure}[t]
\centering
\centerline{\includegraphics[width=0.9\columnwidth]{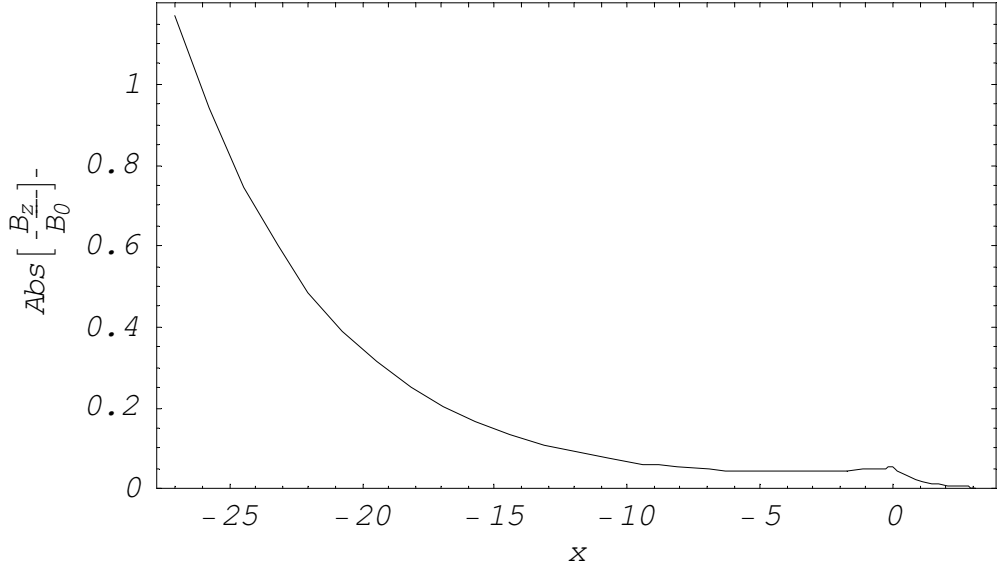}}
\caption{Spatial distribution of an absolute value of the magnetic field
when there is no surface wave present on a plasma-plasma boundary.
Undercritical and overcritical plasma slabs dielectric constants and
thicknesses are $\epsilon_1=0.3428$, $d=27*c/\omega_{p2}$ and $\epsilon_2$%
=-2.97, $a=3.12*c/\omega_{p2}$ correspondingly. External radiation frequency
and angle of incidence are $\omega/\omega_{p2}=0.5019$, $\theta=0.75$
correspondingly. $B_0$ is the magnitude of the magnetic field in the
incident wave.}
\label{fig7}
\end{figure}

As one can see from Figures  (\ref{fig3}) and (\ref{fig4}) the light
transmission is a non-monotonous  function of the widths of both plasma
layers with a single maximum reached at certain  correlated values. These
values are determined by the interference condition between  the two
non-propagating (evanescent) fields.  The evanescent fields are expressed
through the hyperbolic functions  (anomalous transmission occurs when both $%
k_1$ and $k_2$ take on pure imaginary values).  It can be easily seen that
the total field  in both plasma regions constitutes a sum of two exponential
functions, one decaying with distance $e^{-x}$ and the other growing $e^{x}$
($x$ points in the direction from left to right). It is a known fact that a
single evanescent (decaying) wave carries no power. It can be easily shown
that  the $x$ component of the Poynting vector for this wave is $S_x\sim
Re[k]=0$  ($k$ is a purely imaginary spatial decay constant). However, when
the total field  is a superposition of decaying and growing modes, $E,B\sim
A_1\exp[-ikx]+A_2\exp[ikx]$ ($k$ is purely imaginary),  the $x$ component of
the Poynting vector $S_x$ becomes, 
\begin{equation}
S_x=\frac{1}{2}Re[E_yB^*_z]\sim Re[k(A_1A^*_2-A_2A^*_1)].  \nonumber
\end{equation}
Since $A_1A^*_2-A_2A^*_1$ is purely imaginary (it can also be zero when the
incident wave is evanescent and no dissipation is present), $%
Re[k(A_1A^*_2-A_2A^*_1)]\ne 0$, signifying that the superposition of the
decaying and growing modes carry a finite power. We shall call this
superposition as the interference of the evanescent waves. The total light
transmittance of an opaque plasma slab occurs when the Poynting vector
inside the slab is equal to that of the incident radiation, so that $%
Re[k(A_1A^*_2-A_2A^*_1)]=S_{x0}$. Figures (\ref{fig5}) and (\ref{fig6}) show
the spatial distribution of the magnetic field modulus and $x$-component of
time-averaged Poynting vector $S_x=\frac{1}{2}Re[E_yB^*_z]$ for the case
when surface wave is excited on a plasma-plasma interface. As one can see,
the field amplitude grows dramatically in the first plasma layer (the
magnitude of the magnetic field at the interface is more than fourteen times
of that in the incident wave)  to offset its exponential decay in the second
to the level of the incident light.  The $x$-component of the Poynting
vector inside plasma regions is equal to that of the  incident radiation-the
energy of the field is perfectly transferred through the overcritical 
density plasma slab. For comparison, Fig. (\ref{fig7}) shows the magnetic
field amplitude  for the case when there is no surface mode excited at the
interface. Only exponentially  decaying solution to the field equations is
observed. It should be noted here that light  can be transmitted through an
opaque high-density slab even without surface waves  present at the
boundary, if the slab is thin enough. The effect of the evanescent wave 
interference is at play in this case too. The difference lies in the actual
values of  the expansion coefficients $A_1$ and $A_2$ in both cases. When
there is no surface wave  present, the coefficients are such that $%
Re[k(A_1A^*_2-A_2A^*_1)]<S_{x0}$, whereas if the  surface wave is excited at
the interface, the  condition $Re[k(A_1A^*_2-A_2A^*_1)]=S_{x0}$ can be
fulfilled and the absolute  transparency of optically opaque plasma layer is
observed.

\section{Surface waves on a warm plasma-plasma interface}

The results obtained in the previous sections are applicable to the cold
plasma case,  limiting its applications to mostly solid-state plasmas.  In
plasma states created through  the interaction of the high power
electromagnetic radiation with solid structures,  the electron temperature
can reach values in excess of 1 keV. Therefore, it is  important to
understand how the finite temperature modifies the  dispersion relation of
the surface wave and  subsequently the anomalous light transmission through
the dense plasma.

Kaw and McBride\cite{kaw70} have used hydrodynamic description to
investigate the finite temperature dispersion relation for electromagnetic
surface waves propagating  on a plasma-vacuum interface. We generalize their
treatment for the case of surface  waves on a warm plasma-plasma interface.
The dispersion relation is found from solving the linearized set of warm
fluid-Maxwell equations, 
\begin{eqnarray}
\mathbf{\nabla \times E}=-\frac{1}{c}\frac{\partial \mathbf{B}}{\partial t} 
\nonumber \\
\mathbf{\nabla \times B}=-\frac{4\pi n_0 e}{c}\mathbf{\tilde{v}}+\frac{1}{c}%
\frac{\partial \mathbf{E}}{\partial t}  \nonumber \\
\mathbf{\nabla \cdot B}=0  \nonumber \\
\mathbf{\nabla \cdot E}=-4\pi e \tilde{n}  \nonumber \\
\frac{\partial \tilde{n}}{\partial t}+\mathbf{\nabla \cdot} n_0\mathbf{%
\tilde{v}}=0  \nonumber \\
\frac{\partial \mathbf{\tilde{v}}}{\partial t}=-\frac{e}{m}\mathbf{E}-\frac{%
v^2_{Th}}{n_0}\mathbf{\nabla}\tilde{n},  \nonumber
\end{eqnarray}
where the subscript 0 indicates equilibrium quantities; $v_{Th}$ is electron
thermal velocity; ions are assumed to be motionless. The above system of
equations constitutes a closed set for the unknown electron density,
hydrodynamic velocity and electromagnetic fields. Just like in the cold
case, it can be readily solved assuming that density and field perturbations
are in the form $\mathbf{\Psi}(x)\exp[i(k_y y-\omega t)]$, where $x$ is the
coordinate into the plasma and $y$ runs along the interface. This gives the
following expressions for the electromagnetic fields and electron density
perturbation in both plasma regions assuming that the interface is at $x=0$
and omitting the explicit dependence on $\omega$ and $k_y$, 
\begin{subequations}
\label{eqs7}
\begin{eqnarray}
\mathbf{E}_{(1)}=\mathbf{D}_{(1)}\exp[\alpha_1x]-\frac{4\pi
eA(1-\beta^2_1)}{\gamma^2_1-\alpha^2_1}\cdot
\left(ik_y\mathbf{\hat{e}_y}+\gamma_1\mathbf{\hat{e}_x}\right)\exp[\gamma_1x], & x<0\label{eq71}\\
\mathbf{E}_{(2)}=\mathbf{D}_{(2)}\exp[-\alpha_2x]-\frac{4\pi
eB(1-\beta^2_2)}{\gamma^2_2-\alpha^2_2}\cdot
\left(ik_y\mathbf{\hat{e}_y}-\gamma_2\mathbf{\hat{e}_x}\right)\exp[-\gamma_2x],
& x>0\label{eq72} \\
\tilde{n}_{(1)}=A \exp[\gamma_1 x], & x<0\label{eq73}\\
\tilde{n}_{(2)}=B \exp[-\gamma_2 x], & x>0\label{eq74}
\end{eqnarray}
\end{subequations}
where $A$, $B$, $\mathbf{D}$ are unknown coefficients, $\beta=v_{Th}/c$, $%
\mathbf{\hat{e}_y}$, $\mathbf{\hat{e}_x}$ are the unit vectors in $y$ and $x$
directions respectively, $\gamma=\left(k^2_y+\frac{\omega^2_{p}-\omega^2}{%
v^2_{Th}}\right)^{1/2}$, $\alpha=\left(k^2_y+\frac{\omega^2_{p}-\omega^2}{c^2%
}\right)^{1/2}$; subscripts $1,2$ designate both regions. As one can see the
electromagnetic field in a bulk plasma is represented as a superposition of
a transverse and longitudinal perturbations that are completely decoupled
from each other. At this point we would like to bring reader's attention to
the fact that just as in the cold plasma case, solution (\ref{eqs7})
describes decaying branches for the transverse fields as well as for a
longitudinal field in the overcritical plasma region (described by the decay
factor $\gamma_2$ in our notations). However, the density perturbation as
well as the associated longitudinal electric field in the undercritical
plasma region are described not by an evanescent/decaying branch, but rather
by a propagating oscillatory function ($\gamma_1$ takes on purely imaginary
value in this region). The presence of non-decaying oscillations in the
longitudinal field signifies the existence of a bulk plasma wave in the
undercritical plasma region. The existence of the bulk plasmon stems from
the linear wave conversion process that occurs at the plasma boundary\cite
{kondratenko}. In the conversion process, the surface plasma wave gives rise
to a bulk longitudinal plasma mode that propagates away from the interface
into the undercritical density plasma.

The dispersion relation for surface plasma waves on a warm plasma-plasma
interface is found from the boundary conditions on the electromagnetic
fields (tangential components of both fields e.q. $E_y$ and $B_z$ should be
continuous across the interface). The two additional boundary conditions are
obtained from the integration of the continuity equation and the equation of
motion over $x$ coordinate in the limits $x=\pm \delta$ and setting
subsequently $\delta=0$. Assuming that the density perturbation does not
have any singularities at the interface (no surface charge present), the two
additional boundary conditions have the following form, 
\begin{eqnarray}
\frac{\tilde{n}}{n_0}|_1=\frac{\tilde{n}}{n_0}|_2  \nonumber \\
n_0\tilde{V}|_1=n_0\tilde{V}|_2  \label{flux}
\end{eqnarray}
These are the continuity conditions for the electron density (normalized to
its corresponding equilibrium value) and flux perturbations across the
interface. In order to solve for all six unknowns in the system, the above
four boundary conditions have to be also supplemented by the requirement
that the divergence of the transverse part of the total electric field given
by Eqs.~(\ref{eqs7}) be equal to zero. Solving the system of six algebraic
equations, one obtains 
\begin{eqnarray}
&\left[\left(\omega^2-\Delta\right)\sqrt{1-\omega^2+k^2_y}%
+\left(\omega^2-1\right)\sqrt{\Delta-\omega^2+k^2_y}\right]  \nonumber \\
& \times\left[\left(\omega^2-\Delta\right)\sqrt{\frac{1-\omega^2}{\beta^2}%
+k^2_y}+\Delta \left(\omega^2- 1\right)\sqrt{\frac{\Delta-\omega^2}{\beta^2}%
+k^2_y}\right]=\left(\Delta-1\right)^2\omega^2k^2_y ,  \label{eqs8}
\end{eqnarray}
where $\Delta=\omega^2_{p1}/\omega^2_{p2}$, $\omega$ is normalized to the
plasma frequency in the region 2 where $\epsilon<0$ and $k_y$ is normalized
to the classical skin depth $\delta=c/\omega_{p2}$. Eq.~(\ref{eqs8}) is the
general dispersion relation for surface plasma waves running on a warm
homogeneous plasma-plasma interface. It is interesting to note that matching
electromagnetic fields across the interface couples their longitudinal and
transverse components, so that the resulting surface plasma mode is of mixed
longitudinal-transverse type. Taking the limit $\Delta\rightarrow 0$ reduces
this dispersion relation to that derived by Kaw~\cite{kaw70} for the case of
plasma-vacuum interface. The cold plasma dispersion relation is recovered
from the limit $\beta\rightarrow 0$. Numerical solution of Eq.~(\ref{eqs8})
shows that for $\beta_{1,2}\apprle$ 0.02, which corresponds to electron
temperatures 204 eV or lower, the cold plasma dispersion relation (\ref{eqs3}%
) yields solutions that are close to those given by the general expression (%
\ref{eqs8}), neglecting small imaginary parts in the eigenroots for a case
of a warm plasma. The presence of imaginary values in the dispersion
relation is directly linked to already mentioned conversion process. Because
of the surface wave transformation into a bulk plasmon, its energy leaks out
of the surface mode into a bulk plasma mode, which can be viewed as an
effective dissipation of the surface plasmon, making the dispersion relation
complex. It should be noted that the complex conjugate of the roots (with
simultaneous reversal of the real parts of $\omega\rightarrow-\omega$ and $%
k_y\rightarrow -k_y$) are also solutions to the dispersion relation (\ref
{eqs8}), which is physically related to the opposite process in which the
bulk plasmon converts into the surface wave.

\begin{figure}[t]
\centering
\centerline{\includegraphics[width=0.9\columnwidth]{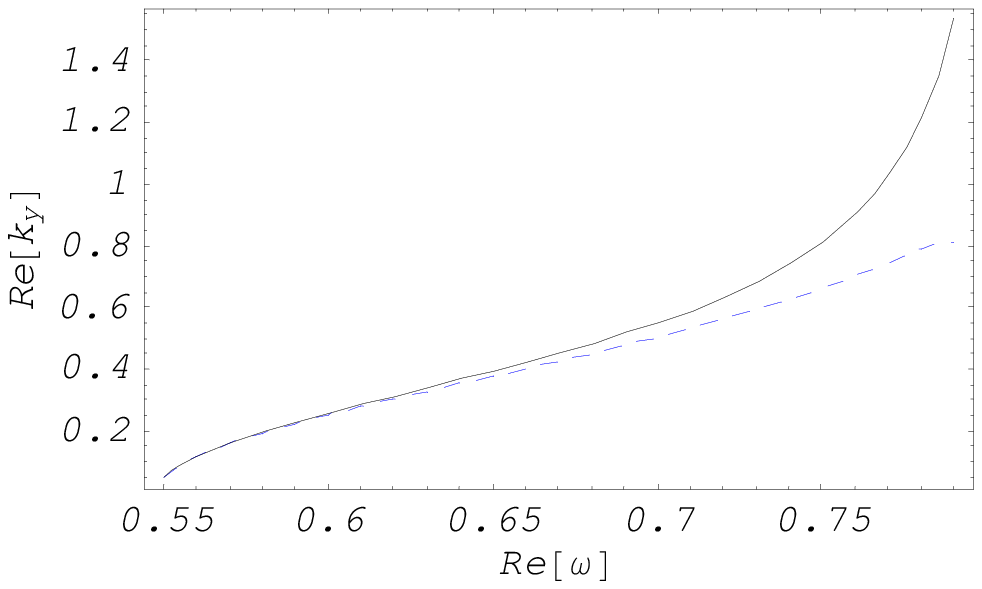}}
\caption{Dispersion relation for surface waves on plasma-plasma interface.
The dashed line is the numerical solution of Eq. (\ref{eqs3}) for $\Delta=0.3
$, $\beta_1=\beta_2=0.1$. The solid line corresponds to the cold plasma
limit with $\Delta=0.3$. $\omega$ is normalized to the plasma frequency of
an overcritical density region and $k_y$ is normalized to its classical skin
depth $c/\omega_{p2}$}
\label{fig8}
\end{figure}

As plasma temperature gets higher, the eigenroots obtained from Eq.~(\ref
{eqs8}) can significantly deviate from those given in the cold plasma limit
as can be seen from Fig. (\ref{fig8}). This fact should be taken into
consideration when investigating a possible surface wave induced plasma
heating channel in ICF experiments where the initial temperatures (due to
slab interaction with laser prepulse) can exceed 1 keV for ultra-intense
laser pulses.

\section{Temperature effects and the anomalous light transmission}

Let us consider a structure similar to that of the cold plasma case and
characterized  by the spatial density (dielectric constant) distribution
shown in Fig.~(\ref{fig1}), but with non-zero electron temperature.  A $p$
-polarized plane electromagnetic wave is obliquely incident on this
structure.  In both plasma regions, the electromagnetic fields $\mathbf{B}%
=(0,0,B_z)$, $\mathbf{E}=(E_x,E_y,0)$ are the solution to the following
system of equations: 
\begin{subequations}
\label{eqs9}
\begin{eqnarray}
\beta^2\frac{d^2E_x}{dx^2}+\left(\frac{\omega^2}{c^2}\epsilon -k^2_y\right)E_x=ik_y\left(1-\beta^2 \right)\frac{dE_y}{dx}\label{eq91}\\
\frac{d^2E_y}{dx^2}+\left(\frac{\omega^2}{c^2}\epsilon -\beta^2 k^2_y\right)E_y=ik_y\left(1-\beta^2 \right)\frac{dE_x}{dx}\label{eq92}\\
\epsilon \frac{d}{dx}\left(\frac{1}{\epsilon}\frac{dB_z}{dx}\right)+\frac{\omega^2}{c^2}\left(\epsilon - \sin^2\theta\right)B_z=0,\label{eq93}
\end{eqnarray}
\end{subequations}
where $k_y=\omega/c$ $\sin~\theta$ and $\beta^2=3v^2_{Th}/c^2$ (the factor
of 3 originates from the adiabatic equation of state) and $%
\epsilon=1-\omega^2_{p}/\omega^2$. The system of equations~(\ref{eqs9}) can
be readily obtained from the initial set of Maxwell and warm fluid
equations. It should be emphasized that Eqs.~(\ref{eqs9}) are only valid for
not too high electron temperatures, so that the electron mean free path is
smaller than the spatial inhomogeneity of the field inside the plasma.
Otherwise, the hydrodynamic treatment is not adequate and one has to invoke
kinetic description, as in a problem of anomalous skin effect\cite
{ichimaru,dragila91}. A general solution to the above system is 
\begin{subequations}
\label{eqs10}
\begin{eqnarray}
B_z=A_1e^{ikx}+A_2e^{-ikx}\label{eq101}\\
E_x=-\frac{k_y}{k_0\epsilon}\left(A_1e^{ikx}+A_2e^{-ikx}\right)-\frac{k_0}{k_y}\left(C_1e^{ik_px}+C_2e^{-ik_px}\right) \label{eq102}\\
E_y=\frac{k}{k_0\epsilon}\left(A_1e^{ikx}-A_2e^{-ikx}\right)-\frac{k_0}{k_p}\left(C_1e^{ik_px}-C_2e^{-ik_px}\right), \label{eq103}
\end{eqnarray}
\end{subequations}
where $k_0=\omega/c$, $k=k_0\sqrt{(\epsilon-\sin^2~\theta)}$, $k_p=k_0\sqrt{%
(\epsilon-\beta^2\sin^2~\theta)}/\beta$. Coefficients $A_{1,2}$ represent
transverse part of the total field and $C_{1,2}$ represent its
longitudinal/electrostatic part. The electromagnetic fields in the vacuum
regions ($x<-d$ and $x>a$) have the following form, 
\begin{eqnarray}
B_z=E_0e^{ik_0\cos \theta~x}+Re^{-ik_0\cos \theta~x},& x<-d  \nonumber \\
E_x=-E_0\sin \theta e^{ik_0\cos \theta~x}-R\sin \theta e^{-ik_0\cos
\theta~x}, & x<-d  \nonumber \\
E_y=E_0\cos \theta e^{ik_0\cos \theta~x}-R\cos \theta e^{-ik_0\cos
\theta~x}, & x<-d  \nonumber \\
B_z=Te^{ik_0\cos \theta~x},& x>a  \nonumber \\
E_x=-T\sin \theta e^{ik_0\cos \theta~x}, & x>a  \nonumber \\
E_y=T\cos \theta e^{ik_0\cos \theta~x}, & x>a  \nonumber \\
\nonumber
\end{eqnarray}
where $T$ and $R$ designate the field components for the transmitted and
reflected waves. By matching the solutions at all boundaries (requiring
continuity of $E_y$, $B_z$, $n_0\tilde{V}_x$ and $\tilde{n}/n_0$ across
interfaces), one obtains a system of ten equations (not shown here) for the
unknown coefficients. It should be noted that in Ref.~\cite{dragila87} the
authors used similar approach to investigate the anomalous transmission of
light through a structure consisting of an overdense warm plasma slab spaced
between two underdense plasma layers. However, the choice of boundary
conditions on the fields used by the authors, specifically the continuity of
the spatial derivative of the $x$-component ($z$-component in their
notations) of the electric field along $x$ leads to non-physical solutions
in which the transmission $T=\left|\frac{T}{E_0}\right|^2 $ or reflection $%
R=\left|\frac{R}{E_0}\right|^2$ coefficients can be greater than one.
Boundary conditions employed in current work  lead to solutions that always
satisfy the identity $|R|^2+|T|^2=1$ (which is the  conservation of energy
for a plain-wave case for a non-dissipative system).  The resulting
expressions for the transmission/reflection coefficients are  extremely
bulky functions of electron temperature, electromagnetic wave incidence 
angle, its frequency, both plasma layers electron densities and thicknesses.

\begin{figure}[t]
\centering
\centerline{\includegraphics[width=0.9\columnwidth]{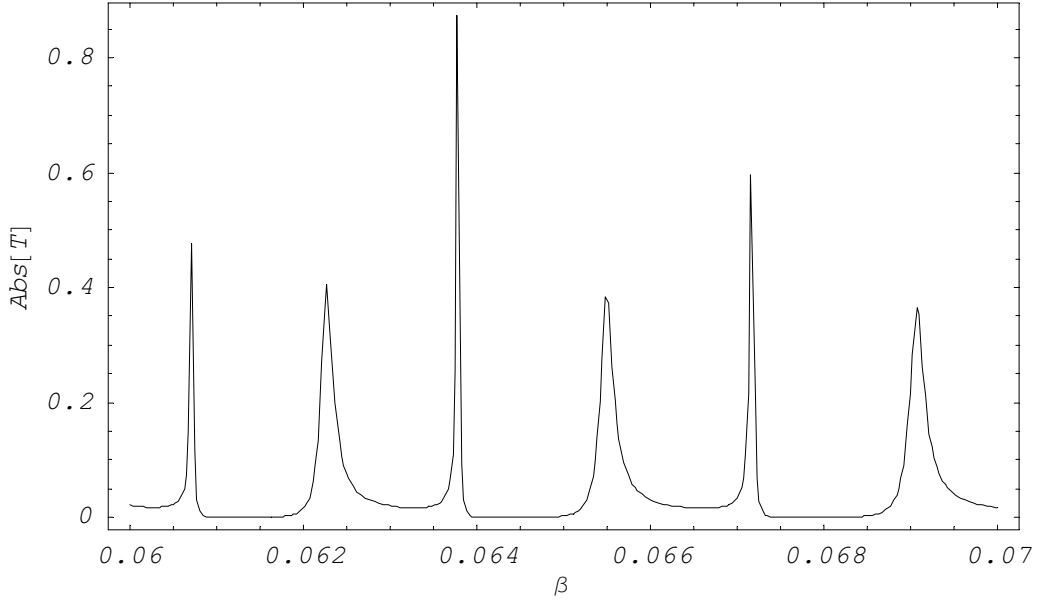}}
\caption{Transmission coefficient as a function of electron temperature $%
\beta=\protect\sqrt{3T/m_ec^2}$. Undercritical and overcritical plasma slabs
dielectric constants and thicknesses are $\epsilon_1=0.3428$, $%
d=27*c/\omega_{p2}$ and $\epsilon_2$=-2.97, $a=3.12*c/\omega_{p2}$
correspondingly. External radiation frequency and angle of incidence are $%
\omega/\omega_{p2}=0.5019$, $\theta=0.671955$ correspondingly.}
\label{fig9}
\end{figure}

As mentioned in the previous section, the dispersion relation for  the
surface plasma waves at non-zero temperatures can significantly  deviate
from that obtained in the cold plasma limit. Since the anomalous
transmission  of light is due to the excitation of the surface plasmons, the
resulting transmission  coefficient exhibits thermal effects that are not
seen in cold plasmas.  Figure (\ref{fig9}) shows the dependence of the
transmission coefficient on the thermal  electron velocity $v_{Th}=\sqrt{%
T/m_e}$. As one can see, it shows high frequency  oscillations in the
electron temperature space. The presence of these regular oscillations  in
the transmission of the system is related to the linear wave conversion
processes  that occur at the plasma boundaries. The surface wave excites a
bulk longitudinal plasma  mode  propagating in the underdense plasma layer
away from the plasma-plasma boundary.  When it reaches the vacuum-plasma
interface, the longitudinal plasma wave  converts into a transverse
propagating electromagnetic wave, which  interferes with the reflected
light. Thus, in the general case the interference  between the
incident/reflected fields, surface plasma wave induced fields  and the
fields created through the linear conversion processes determine  the
transmission.

\begin{figure}[t]
\centering
\centerline{\includegraphics[width=0.9\columnwidth]{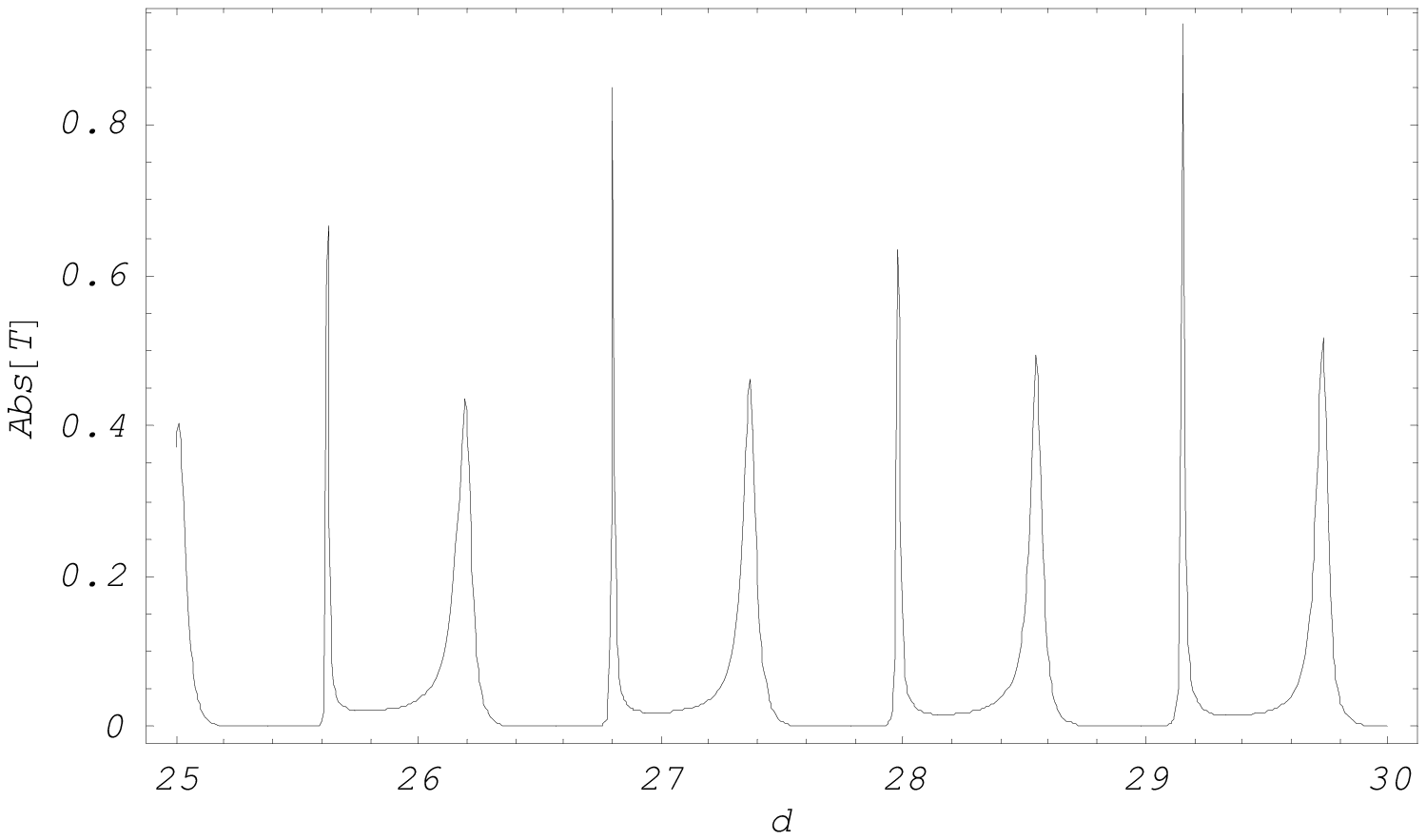}}
\caption{Transmission coefficient as a function of a underdense plasma layer
thickness $d$ for plasma temperature parameter $\beta=0.055$. Overcritical
plasma slab dielectric constant and thickness are $\epsilon_2$=-2.97, $%
a=3.12*c/\omega_{p2}$ correspondingly. Underdense plasma layer dielectric
constant is $\epsilon_1=0.3428$. External radiation frequency $%
\omega/\omega_{p2}=0.5019$. Incidence angle $\theta=0.671955$. }
\label{fig10}
\end{figure}

\begin{figure}[t]
\centering
\centerline{\includegraphics[width=0.9\columnwidth]{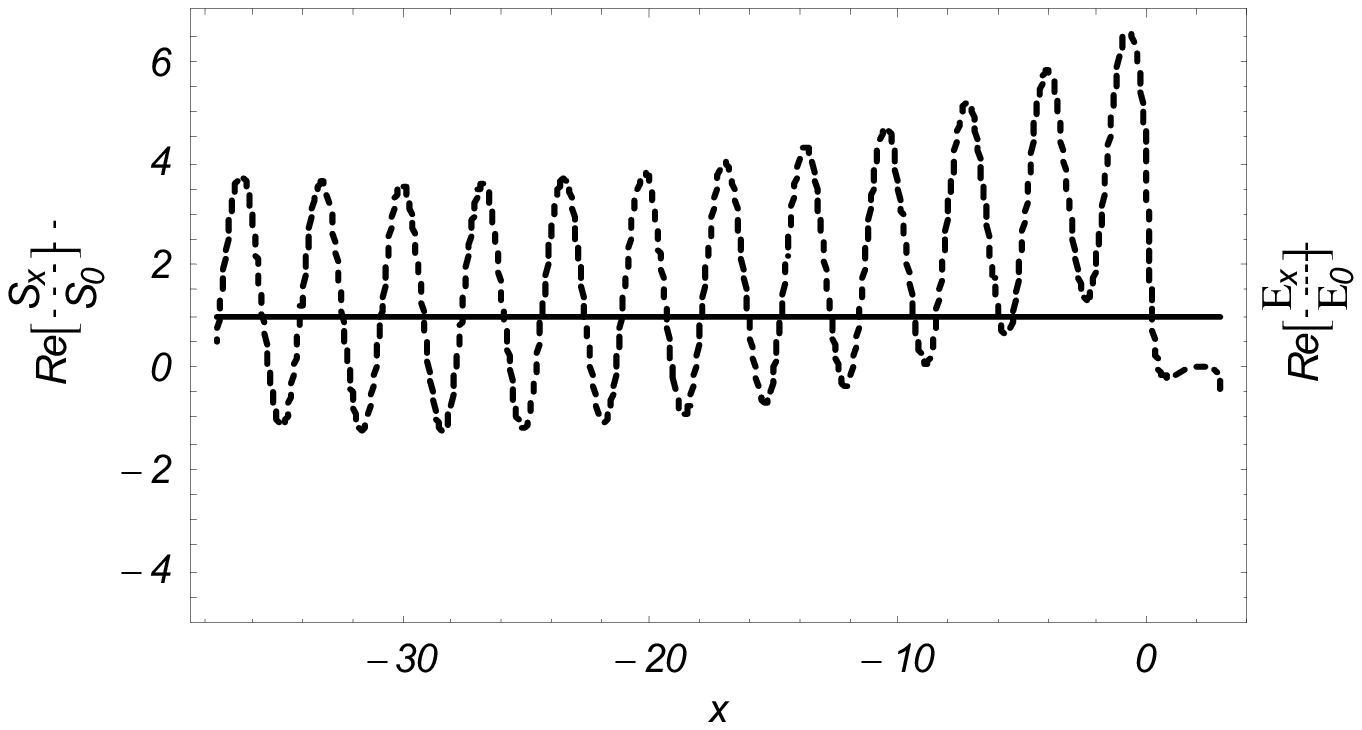}}
\caption{Spatial distributions of the $x$-components of the Poynting vector $%
S_x$ (solid line) and the total electric field (dashed line) inside plasma
layers for plasma temperature parameter $\beta=0.15$. Overcritical plasma
slab dielectric constant and thickness are $\epsilon_2$=-2.97, $%
a=3.*c/\omega_{p2}$ correspondingly. Underdense plasma layer dielectric
constant and thickness are $\epsilon_1=0.3428$, $d=37.5*c/\omega_{p2}$ .
External radiation frequency $\omega/\omega_{p2}=0.5019$. Incidence angle $%
\theta=0.64516$.}
\label{fig11}
\end{figure}

The temperature oscillations can be  understood using the following
simplified model. The surface plasma wave  excites a bulk plasmon
propagating at angle $\theta ^{\prime}$ to the surface normal.  This angle
can be found from matching the $y$ components of both modes' wave  numbers
to give: 
\begin{equation}
\sin\theta ^{\prime}=\frac{\beta}{\sqrt{\epsilon}}\sin\theta
\end{equation}
The plasma wave propagates the distance $L=d/\cos\theta ^{\prime}$ with
phase velocity $v_{ph}=c\beta/\sqrt{\epsilon}$. Therefore, the phase
difference $\Delta \phi=\omega\Delta t$ between the reflected light and that
emitted through the conversion process is, 
\begin{equation}  \label{eqs11}
\Delta \phi=\frac{k_0d\sqrt{\frac{\epsilon}{\beta^2}}}{\sqrt{1-\frac{\beta^2%
}{\epsilon}\sin^2\theta}}.
\end{equation}
When the phase difference is equal to a multiple of $2\pi$, both components
interfere constructively giving rise to large reflection and thus low
transmission. Therefore, one arrives at a following temperature condition at
which minimal light transmission occurs through the system, 
\begin{eqnarray}  \label{eqs12}
\beta=\sqrt{\frac{1}{2}\epsilon\csc^2\theta-\frac{\epsilon\csc^2\theta \sqrt{%
\pi^2n^2-d^2k^2_0\sin^2\theta}}{2\pi n}}, \ n=1,2,3,.... \\
\nonumber
\end{eqnarray}
With a simple estimate given by Eq.~(\ref{eqs12}), we are able to recover
some of the minima in the transmission coefficient obtained from the
complete solution of Eqs.~(\ref{eqs8}). Other minima in the transmission
coefficient, which are not described by Eq.~(\ref{eqs12}) are due to other
conversion processes like the incident wave transformation into a
longitudinal plasma mode at the vacuum-plasma interface that travels to the
plasma-plasma interface, gets reflected (exciting a surface mode in a
process) and travels back to the plasma-vacuum interface converting to the
transverse electromagnetic wave. All these conversion processes contribute
to the transmission properties of the double-plasma system, making it highly
oscillatory function of not only the electron temperature but also the
underdense plasma layer thickness $d$ as can be seem from Fig. (\ref{fig10}%
). The multiple peaks present in the transmission coefficient in this case
can also be explained through the constructive interference between the
reflected electromagnetic field and the field created as a result of the
linear wave conversion process. Since the phase difference $\Delta \phi$
also depends on the underdense layer thickness, one can arrive at a similar
condition for the plasma thickness $d$ at which minimal light transmission
occurs through the system, 
\begin{eqnarray}  \label{eqs13}
d=\frac{2\pi n\sqrt{1-\frac{\beta^2\sin^2\theta}{\epsilon}}}{k_0\sqrt{\frac{%
\epsilon} {\beta^2}}}, \ n=1,2,3,.... \\
\nonumber
\end{eqnarray}
A simple estimate given by Eq.~(\ref{eqs13}) can also recover the minima in
the transmission coefficient as a function of the undercritical density
plasma layer thickness for a case of non-zero electron temperature. As the
plasma temperature increases, the resonant condition for the excitation of
the surface wave changes according to the dispersion relation given by Eq.~(%
\ref{eqs8}), shifting the peak in the transmission coefficient to smaller
angles. A distinct difference between the zero temperature and a warm plasma
case lies in a fact that there is only a single peak in the transmission
coefficient (as a function of the undercritical density plasma layer
thickness) in the cold plasma case, whereas the number of transmission peaks
in a warm plasma is quite large. Unlike the cold plasma case where the
energy of the external radiation is transferred through the constructive
interference of evanescent fields, in a warm plasma case the superposition
of the propagating longitudinal plasma wave fields (in a undercritical
density plasma slab) and the evanescent transverse electromagnetic fields
determine the amount of the transmitted light as can be seen from figure (%
\ref{fig11}), where the $x$ components of time-averaged total (which
includes the contribution of plasma electrons) Poynting vector and the real
part of the electric field are shown. As expected, the electromagnetic field
has a form of the superposition of the evanescent and
propagating/oscillatory branches and under resonant condition the energy is
totally transferred through the system. In addition, the results of our
calculations show that in a warm underdense plasma layer the energy is
mainly carried by the longitudinal electron plasma oscillations,
facilitating the total transparency of the considered dual-layer system for
quite large thicknesses of an undercritical density plasma slab.

\section{conclusions}

It is shown here that overcritical density plasma can be made transparent to
external radiation. The total transparency can be achieved even with a
single surface mode excited on a plasma-plasma boundary. The anomalous
transmission is explained through the interference effect between the
evanescent fields of the incident electromagnetic radiation and those
associated with the surface plasmons. Superposition of two (decaying and
growing)  evanescent modes can carry the power of the incident radiation
over to the other side of a high-density plasma slab, making it totally
transparent to light in a case when irreversible dissipation is absent. We
have also shown that the excitation of the longitudinal plasma waves creates
a multiplicity of additional resonances corresponding to the absolute
transparency. We suggest that the mechanism investigated in this paper may
be responsible  for the anomalous transparency of metals observed recently
in a number of experiments \cite{pendry00,fang05}.

The role of dissipation effects is twofold, depending on possible
applications of the surface wave phenomenon. In ICF experiments where high
absorption of the incident radiation is sought, the presence of irreversible
dissipation is necessary for effective heating of a fuel pellet through
damping of the surface plasma waves. If dissipation is absent, the energy of
the incident light will be totally re-emitted by these waves back into
vacuum. If on the other hand a high transmission of light is needed, the
dissipation plays a detrimental role, limiting the transmitting ability of
the system. Therefore, a close control over a number of physical parameters
(electron density, temperature, thickness) of the double-layer system is
needed for successful technological implementation of the surface wave
effect.

The final point we would like to mention is related to the fact that since
the Maxwell equations are invariant under reversal of time and space
coordinates, the order at which the incident electromagnetic wave encounters
both plasma layers (an undercritical density layer followed by that with
overcritical density or vice versa) does not effect its transmission ability.

\acknowledgements

This work is in part supported by NIH(CA78331), Strawbridge Family
Foundation, Varian medical systems and NSERC Canada.

\bibliographystyle{plain}

\end{document}